\documentclass[twocolumn,final,epj]{svjour}
\usepackage[utf8]{inputenc}
\usepackage[T1]{fontenc} 
\usepackage{graphicx}  
\usepackage{amssymb}
\usepackage{amsmath}
\usepackage{hyperref}
\hypersetup{
    colorlinks=true,       % false: boxed links; true: colored links
    linkcolor=blue,          % color of internal links (change box color with linkbordercolor)
    citecolor=blue,        % color of links to bibliography
    filecolor=magenta,      % color of file links
    urlcolor=cyan           % color of external links
}
\usepackage{color}

\begin{document}

\title{Opinion formation at Ising social networks}

\author{
Kristina Bukina\inst{1}${^{(*)}}$ 
\and
Dima L. Shepelyansky\inst{1}}

\institute{
\inst{1} Univ. Toulouse, CNRS, Laboratoire de Physique Th\'eorique, Toulouse, France
}

\titlerunning{Opinion formation at Ising social networks}
\authorrunning{K.Bukina and D.L.~Shepelyansky}

%\date{\today}
\date{Dated: 16 November 2025}

\abstract{We study the process of opinion formation
in an Ising social network of
scientific collaborations. The network is undirected.
An Ising spin is associated with each network node
being oriented up (red) or down (blue).
Certain nodes carry fixed, opposite opinions
whose influence propagates over the other
spins, which are flipped according to
the majority-influence opinion of neighbors of a given spin
during the asynchronous Monte Carlo process.
The amplitude influence of each spin
is self-consistently adapted,
and a flip occurs only
if this majority influence exceeds a certain
conviction threshold.
All non-fixed spins are initially
randomly distributed, with half of them oriented up and
half down. Such a system can be viewed
as a model of elite influence, coming from the fixed spins,
on the opinions of the crowd of non-fixed spins.
We show that a phase transition occurs as the amplitude influence
of the crowd spins increases:
the dominant opinion shifts from that of the elite nodes
to a phase in which the crowd spins' opinion becomes
dominant and the elite can no longer impose
their views.
}

%% PACS to be updated or removed
%\PACS{
%{89.75.Fb}{
%Structures and organization in complex systems}
%\and
%{89.75.Hc}{
%Networks and genealogical trees}
%\and
%{89.20.Hh}{
%World Wide Web, Internet}
%}

\maketitle

\section{Introduction}
\label{sec1}

Social networks now exert a significant influence on human society,
and as a result, their properties are actively investigated
by the scientific community (see e.g.,  \cite{fortunato09,dorogovtsev10,newmanbook}).
Recently their impact has been argued to extend specifically to opinion formation
and even to affect political elections \cite{soc1,soc2}.
This very problem of opinion formation in a group of electors is
actively investigated in the field of sociophysics,
using diverse models and methods
(see e.g. \cite{galam82,galam86,sznajd00,sood05,watts07,galam08,kandiah12,eom15}).
Usually in these studies there are two competing opinions
of electors, often modeled as network nodes,
governed by a local majority rule whereby
an elector's opinion is determined by
the majority opinion of its linked neighbors.
Thus, each node has red or blue color (or an Ising spin up or down),
and the system represents an Ising network of spin halves
with $N$ nodes and a huge space of $N_{conf}=2^N$ configuration states
(see e.g., \cite{galam08}).
An opinion, or spin polarization, of nodes is determined
by an asynchronous Monte Carlo process
in a system of spins described by an Ising Hamiltonian
on a network. A similar Monte Carlo process is used in
the models of associative memory \cite{memory1,memory2}.

Recently it was proposed that such an opinion formation process
can also describe a country's preference to trade 
in one currency or another (e.g. USD or hypothetical BRICS currency) 
\cite{brics}. An important new element  introduced  in \cite{brics},
and then extended in \cite{inof,ising25},
is that the opinion of certain network nodes is considered to be
fixed (spin always up or down) and not affected by opinions of other nodes.
In addition, in such an Ising Network of Opinion Formation (INOF) model
\cite{inof,ising25} it is assumed that at the initial stage
only fixed nodes have a given fixed spin polarization,
while all other nodes are white (zero spin)
thus producing no influence on the opinions (spins)
of other nodes. However, these white nodes
are getting their spin polarization up or down
during the asynchronous Monte Carlo process
of opinion formation on the Ising network.
All the above studies  
have been done for directed networks
with the INOF approach of fixed and white nodes
applied to Wikipedia Ising Networks (WIN) 
considering contests between different social concepts \cite{inof},
companies, political leaders and countries \cite{ising25}.
When we consider a contest between two political leaders like Trump and Putin
in WIN, it is rather natural to assume that all other nodes (Wikipedia articles)
have no specific opinion on these two figures
at the initial stage of the Monte Carlo process of INOF,
so that they are considered as white nodes.
However, it may be important to understand the influence of initial random opinions
of non-fixed nodes on the contest results. Beyond this, the INOF approach can be applied
to social networks, which in many
cases are undirected, such as Facebook.
We note that the properties of the Ising model
on complex networks were studied previously
(see e.g. \cite{dorogovtsev,bianconi}),
but the opinion formation process was not studied there.

To this end, in this work we apply the INOF approach
to a social network of scientists
studied by Newman \cite{newman2001,newman2006} with data sets 
from his database \cite{newmannets,newman2006ref84}.
On the basis of this undirected network we study
the process and features of opinion formation
and analyze the effects of randomized opinions of non-fixed
nodes on this process.

The paper is organized as follows:
In Section 2 we describe the data sets and the Generalized INOF (GINOF) model;
Section 3 presents the results, starting with the original INOF model and
then analyzing the phase transition in the GINOF model;
a discussion of the results and conclusions are provided in Section 4.
Certain data sets are also  available at \cite{ourwebpage}.

\section{Data sets and model description}
\label{sec2}

For our studies we choose the social collaborative network of $N=379$ scientists
(nodes), analyzed in \cite{newman2001,newman2006}, taken from
\cite{newmannets}. The network image is available in Fig.~8 at \cite{newman2006}
and in \cite{newman2006ref84}, where the network nodes are given with the names
of scientists. This is an undirected network with weighted symmetric adjacency matrix
$A_{ij}=A_{ji}$ with the number of links $N_\ell = 1828$; the weight of links
changes from a minimal $a_{min}=A_{ij}=0.125$ to a maximal $a_{max}=4.225$ value;
there are no isolated communities in this network.
The average number of links per node is $\kappa=N_{\ell}/N \approx 4.8$.
The effects of nonlinear perturbation
and dynamical thermalization in this network were recently studied in \cite{fpu70}.
The full list of network links and node names
is available at \cite{newmannets,newman2006ref84}
and \cite{ourwebpage}.

As in \cite{fpu70}, we construct the Google matrix
of the network defined in a standard way \cite{meyer,fpu70} as
$G_{ij} = \alpha S_{ij} + (1-\alpha)/N$ where $S_{ij}$ is the matrix of Markov transitions
obtained from $A_{ij}$ by normalizing to unity
all matrix elements in each column. We use here the standard value of damping factor
$\alpha=0.85$. There are no dangling nodes in this network.
The PageRank vector $P_i$ is the solution of the equation $GP =\lambda P$ at $\lambda=1$;
its elements are positive and give a probability to find a random surfer on a node $i$ \cite{meyer}.
By ordering all nodes by a decreasing order of $P_i$, we obtain the PageRank index $K$
changing from $K=1$ at the maximal $P(K)$ to $K=379$ at the minimal $P(K)$.
The top 10 PageRank nodes from $K=1$ to $10$ are: Barabasi, Newman,
Sole, Jeong, Pastorsatorras, Boccaletti, Vespignani, Moreno, Kurths, Stauffer \cite{fpu70}.
All links $A_{ij}$ and PageRank indexes with names are available at \cite{ourwebpage}.

The INOF procedure of opinion formation on Ising networks is described in detail in \cite{ising25}.
It assumes that there is a group of fixed red nodes (spin $\sigma_i=1$)
and another group of fixed blue nodes (spin $\sigma_i =-1$); all other nodes
are white ($\sigma_i=0$) at the initial state but can change their spins
to $\pm 1$ during an asynchronous Monte Carlo process.
Compared to the INOF model \cite{ising25},
here we extend the condition of spin flip and the initial state of white nodes.
Thus, to all originally white nodes we attribute
vote power, or amplitude influence, determined by
coefficients $W_i$ which characterize the level
of an elector's conviction regarding the importance of the election
and/or his interest in elections. Initially, all white nodes have the same $W_i =W < 1$.
For fixed nodes we always have $W_i=1$.
Also, all previously white nodes are randomly assigned spins $\sigma_i=1$ or $\sigma_i=-1$.
Thus, for our network we have 188 red and 188 blue nodes
with a random distribution of colors
(1 node remains white due to the odd number of nodes) and
there are also 2 fixed nodes with opposite spins $\sigma = \pm 1$.
With this initial configuration of all node spins,
the spin $i$ flip condition is determined by
accumulated influence of the opinions of
linked nodes $j$:

\begin{figure}
\begin{center}
\includegraphics[width=0.95\columnwidth]{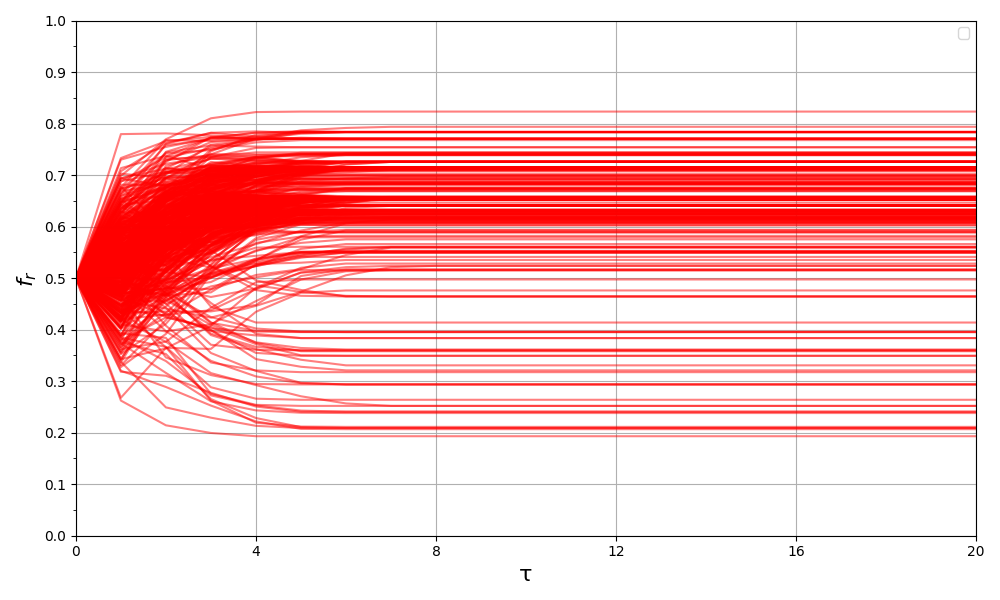}%
\end{center}
\caption{\label{fig1}
Evolution of the fraction of red nodes $f_r$ for $N_r = 500$
random pathway realisations. An initial condition has one 
red fixed node (Newman) and one blue fixed node (Barabasi); they remain
fixed during an asynchronous Monte Carlo evolution
based on the relation (\ref{eqzc});
all other nodes are initially white ($\sigma_j=0$ in (\ref{eqzc})).
Here  $x$-axis represents time time $\tau$ of Monte Carlo process,
where each unit of $\tau$ marks one complete update of all
nodes/spins following the INOF/GINOF model (here $Z_c=0; W=0$);
steady-state configurations are reached at $\tau=20$ (or earlier).
}
\end{figure}

\begin{equation}
\label{eqzc}
Z_i = \sum_{j \neq i} \sigma_j W_j A_{ij}
\end{equation}
Here the sum runs over all $j$ nodes linked to $i$
with the contribution of $A_{ij}$ links and vote power $W_j$.
The flip condition of spin $i$ is defined as:
for $Z_i > Z_c$ its $\sigma_i=1$ and $W_i=1$;
for $Z_i < -Z_c$ its $\sigma_i=-1$ and $W_i=1$;
for $|Z_i| \leq Z_c$ its spin $\sigma_i$ and coefficient $W_i$ remain
unchanged. Thus the parameter $Z_c$ has the meaning of
opinion conviction threshold (OCT)
so that if the module of influence of neighbors $|Z_i|$ is less than $Z_c$, then
the elector $i$ does not take into account their opinions.
Also, if  $|Z_i| > Z_c$, then this elector $i$ becomes convinced
in the importance of this election and
its $W_i =1$ for all future evolution.

This asynchronous Monte Carlo procedure of spin
flips is done for all spins (except fixed ones)
without repetitions. When the run over all spins is done,
we arrive to the Monte Carlo time $\tau=1$,
after that the procedure goes to $\tau =2$
with another random pathway order of spin flips and 
so on till $\tau=20$ when the process converges to
a steady-state. This corresponds to one pathway realisation
for a specific order of spin flips, then the process
is repeated for another pathway realization
of spin flips order and the average
fractions of red $f_r$ and blue $f_b$ nodes (up/down spins)
are determined averaging over all pathway realisations and all nodes,
which gives the total red fraction $f_r$ (by construction
$f_r+f_b=1$ since there are no white nodes in this network
at the steady-state). Several examples of
$\tau-$evolution of red fraction $f_r$ are shown in Fig.~\ref{fig1}.  
We also determine the average fraction of red nodes $f_r(i)$ 
for each node $i$ by averaging over $N_r$ pathway realisations.
We use $N_r=10^4$ and $10^5$ in this work.

We call the INOF model described above
the Generalized INOF model (GINOF). The main new elements of GINOF
are the absence of white nodes at the initial state and
their replacement by non-fixed nodes with a random spin configuration
with half of them spin up and the other half spin down.
However, now each spin of this configuration
has an amplitude influence $W_i<1$ entering
in the influence score $Z_i$ at (\ref{eqzc});
initially all non-fixed nodes have $W_i=W <1$.
A flip of spin $i$ takes place
only if its influence score
exceeds the opinion conviction threshold $Z_c$
with $|Z_i|>Z_c$, in which case
its amplitude influence becomes $W_i=1$
for all further iterations. Evidently,
the fixed nodes always have their $W=1$
and their opinions remain fixed.

In a certain sense, in the GINOF model
the fixed nodes can be viewed as
two competing elite groups with opposite opinions
that try to convince the other members of society
(the crowd of electors) with random
opinions (half red and half blue).
These electors,
at the initial state of election process
have a weak amplitude influence
on the score of other electors ($W<1$).
During the election campaign,
modeled as a Monte Carlo process,
the crowd nodes,
whose influence score
exceeds the opinion conviction threshold $Z_c$,
become active in the election process, acquiring
the maximal amplitude influence $W_i=1$.
For the case with $W_i =W=0$,
the GINOF model is reduced to
the original INOF model studied in \cite{ising25}.

At first glance it seems that
the network with $N=379$ nodes considered here is
much smaller compared to
INOF studies with $N \sim 10^6$ reported in \cite{ising25}.
However, we point out that even with
$N=379$, the number of configuration
states of the Ising network is huge,
being $N_{conf} =2^N$. Also, in the
studies of other spin systems
with an asynchronous Monte Carlo
process, a similar number of nodes
had been considered with $N \approx 400 - 1000$
in \cite{memory2}, and $N \approx 100$ in
\cite{albert,levine}.

The results for the GINOF model are
presented in the next Section.
They show that there is
a transition between two phases:
from a phase where the elite is able to impose its opinion
to a phase where the opinion of the elector crowd
is dominant over the elite opinion.

\section{Results}
\label{sec3}

\subsection{INOF results with white nodes}

We first present the results for the INOF model \cite{inof,ising25}
with initial state, where non-fixed nodes are white.
As the nodes with fixed opinions, we choose
the node of Newman (red, spin up)
and the node of Barabasi (blue, spin down)
(see the network with names of scientists at \cite{newman2006,newman2006ref84}).
We use these two fixed nodes for all other network results
of this work. We point out that such an initial condition
of spin polarization also corresponds to 
the GINOF model at $Z_x=0, W_i=W=0$ as described in the previous section.

\begin{figure}
\begin{center}
\includegraphics[width=0.95\columnwidth]{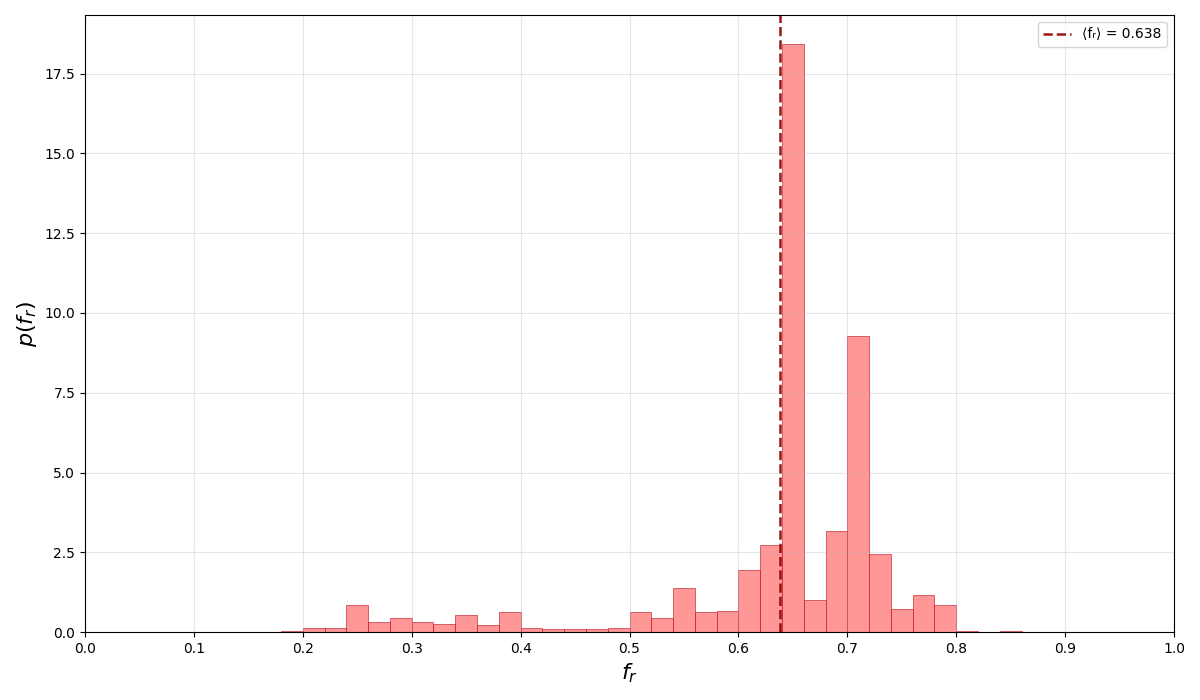}%
\end{center}
\caption{\label{fig2}
  Probability distribution $p(f_r)$ of red node fractions;
  the histogram of $f_r$ values is obtained with 50 cells $1 \leq m \leq 50$
  with normalization $\sum_m f_r(m) =1$,
  average red value is $<f_r> = 0.638$. Here there are  $N_r=10^5$ pathway realizations.
  Fixed nodes are Newman (red) and Barabasi (blue),
  all other nodes are white (spin zero). Initially all non-fixed nodes
  are white for the INOF model  
  [or for the GINOF model at $W=0; Z_c=0$]. Vertical dashed line marks the
  average red value  $<f_r>$.
}
\end{figure}

The histogram of the probability distribution $p(f_r)$
of red fractions $f_r$, obtained in the steady state (at $\tau=20$),
is shown in Fig.~\ref{fig2}. It is obtained by averaging
over $N_r=10^5$ pathway realizations and over all $N=379$ nodes.
The total average fraction of red nodes is
$<f_r>=0.638$, favoring Newman.
The average polarization of all spins is
$\mu_0 = <f_r> - <f_b> =2<f_r>-1 =0.276$.

\begin{figure}
\begin{center}
\includegraphics[width=0.95\columnwidth]{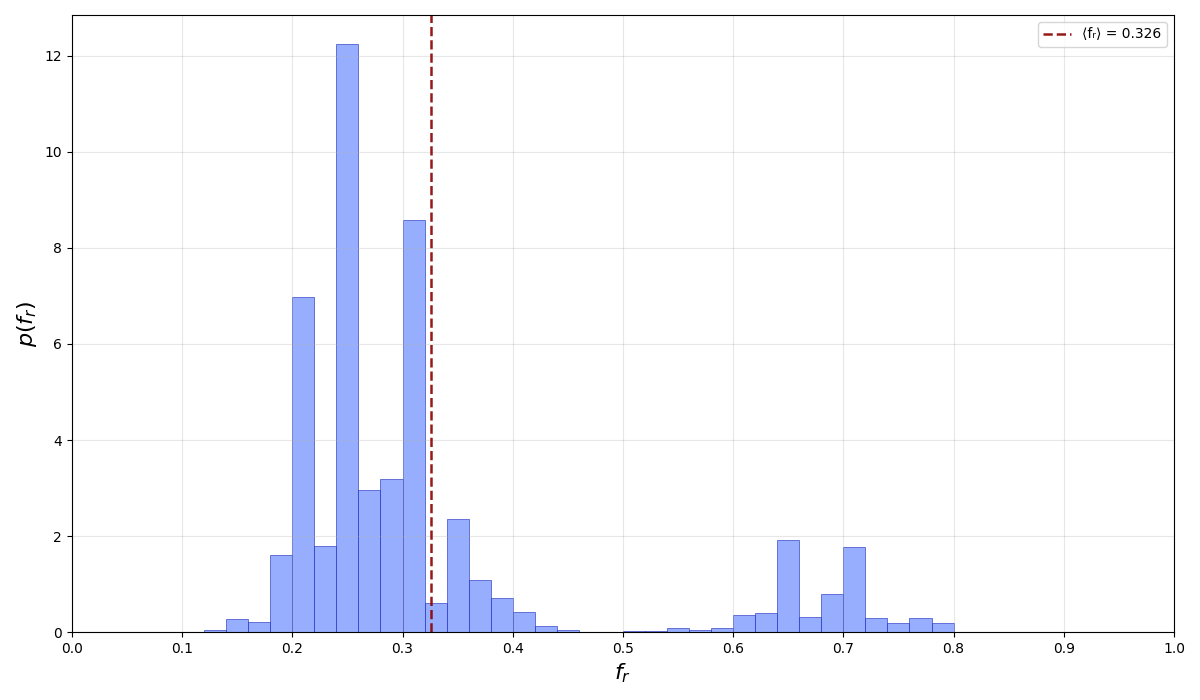}%
\end{center}
\caption{\label{fig3}
  Same as Fig.~\ref{fig2}, but with initial state of node Sole being blue;
   $<f_r> = 0.326$
}
\end{figure}

\begin{figure}
\begin{center}
\includegraphics[width=0.85\columnwidth]{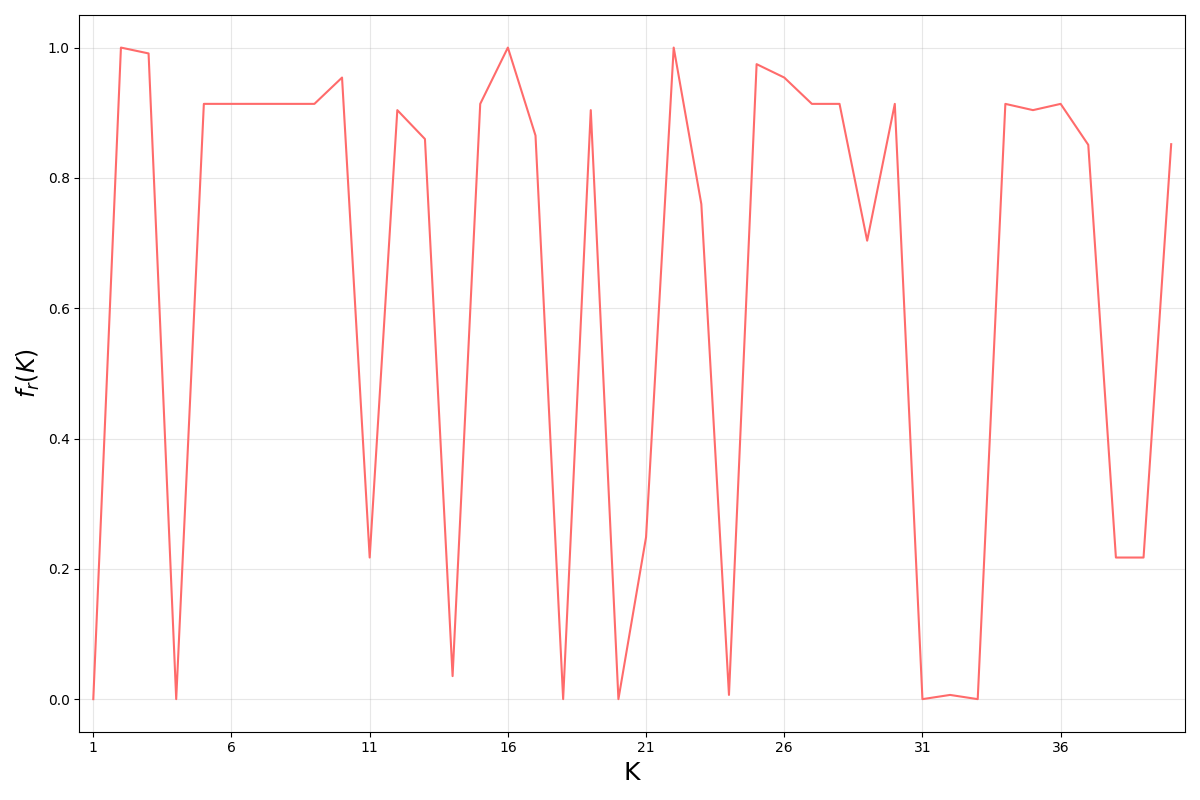}%
\end{center}
\caption{\label{fig4}
  Dependence of the red fraction of nodes $f_r(K)$ on PageRank index $K$ 
  for the case of Fig.~\ref{fig2} ($K$ is obtained at damping factor $\alpha=0.85$).
}
\end{figure}

It is interesting to note that the distribution $p(f_r)$
can be significantly affected if in the initial state
one replaces a certain white node
by initial node with spin up or down (red or blue),
which, however, is not fixed and can be flipped
during the Monte Carlo process.
We show an example of such
a striking influence in Fig.~\ref{fig3},
where the initial white node Sole
(see network with names at \cite{newman2006ref84})
is replaced by a blue node
(all other nodes are the same as in Fig.~\ref{fig2}).
We see that such a one-node change
gives a complete modification of the distribution $p(f_r)$
with the total average probability $<f_r>=0.326$, favoring Barabasi.
The reason for such a strong effect
is the fact
that the Erd\"os number $N_E$ \cite{dorogovtsev10} of
Sole with respect to Newman is $N_E=1$
(direct link between them)
and also that the right part of the whole network
(see \cite{newman2006ref84}) is
linked with Newman mainly via node Sole.
In a certain sense, such a specific
placement of a blue node
in the initial configuration
of colored nodes represents the Erd\"os barrage,
which was also shown to be very efficient
in the case of fibrosis disease propagation
in the MetaCore network of
protein-protein interactions \cite{fibrosis2}.

In the framework of the GINOF model, we obtain not only the average
value of red opinion $<f_r>$ but also the average red opinion
for each node $f_r(K)$, with $K$ being the PageRank index.
The dependence $f_r(K)$ is shown in Fig.~\ref{fig4}
for the top 40 PageRank nodes with $K=1,\cdots, 40$
(all $f_r(K)$ values are available at \cite{ourwebpage}).
For the top 10 PageRank nodes we have
$f_r(K)$ values: 0.000, 1.000, 0.991, 0.000,
0.913, 0.913, 0.913, 0.913, 0.913, 0.954 for
$K=1, \cdots, 10$
(see the corresponding 10 names above). 
Usually the nodes with an Erd\"os number $N_E =1$
with respect to Newman have $f_r =1$ or very close to 1
and similarly for nodes at $N_E=1$ from Barabasi,
with $f_r \approx 0$.
However, there are cases with $N_E=5$
and $f_r(K=9)=0.913$ (Kurths), indicating that the
competition of colors on this social network
has a rather complex structure.
It is also clear that there is no
simple correlation between the top PageRank index and the top values
of the probability of red or blue colors.

\begin{figure}
\begin{center}
\includegraphics[width=0.85\columnwidth]{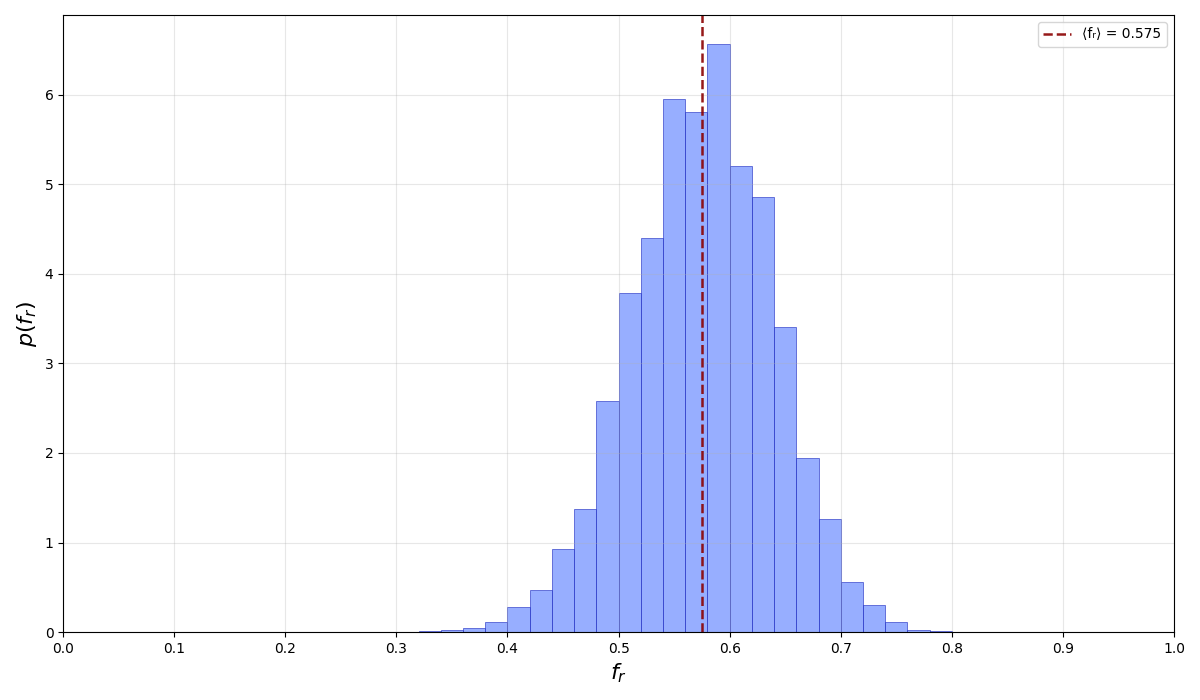}%
\end{center}
\caption{\label{fig5} Same as Fig.~\ref{fig2}
  but for the GINOF model at 
  $Z_c=0$, $W=0.005$; here $N_r = 10^5$.
}
\end{figure}

\subsection{Effects of opinion conviction threshold in GINOF}

One of the important elements of the INOF model
is the presence of white nodes
in the initial state. This can be considered as a natural choice
for Wikipedia and some other directed networks \cite{inof,ising25,fibrosis2}.
However, for the models of election votes on social networks
it may be more consistent to assume that
the elite members of society have fixed opposite opinions
of the leaders of two parties while
the crowd of people (electors)
have random red and blue opinions
with a low initial interest in elections,
and hence a low amplitude influence of their votes $W<1$
(e.g. because only a small fraction of such electors participate in
an election). Thus, we suppose that the GINOF model is more
adequate for modeling elections on social networks.

\begin{figure}
\begin{center}
  \includegraphics[width=0.85\columnwidth]{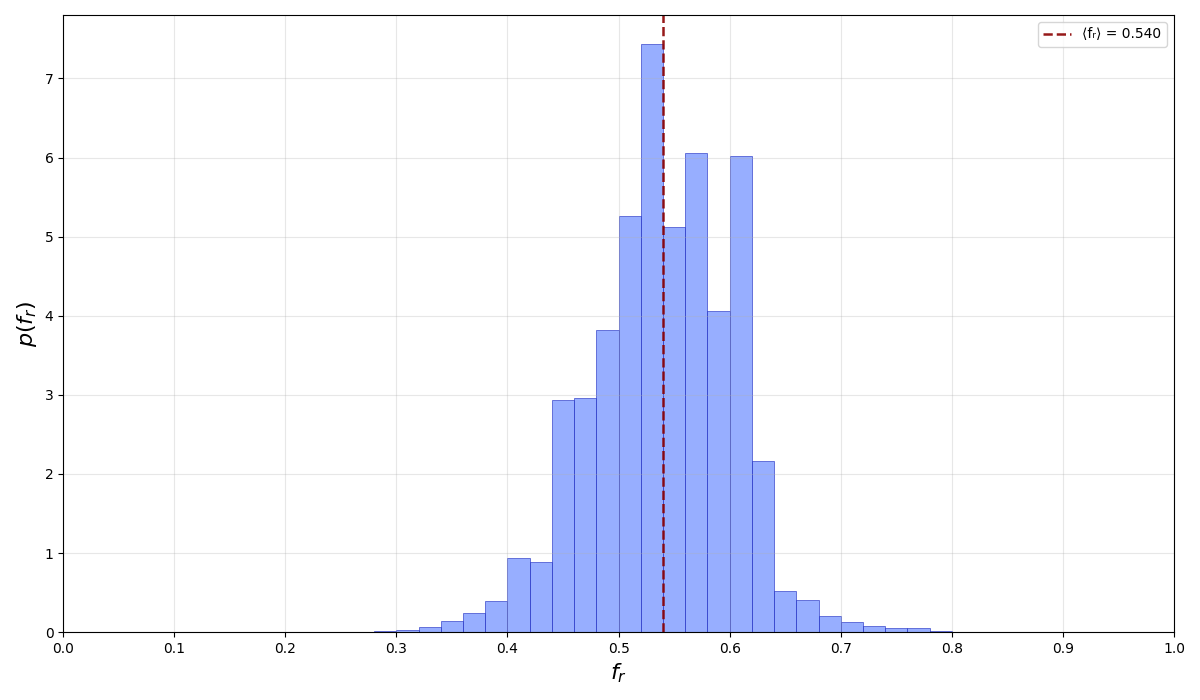}\\
  \includegraphics[width=0.85\columnwidth]{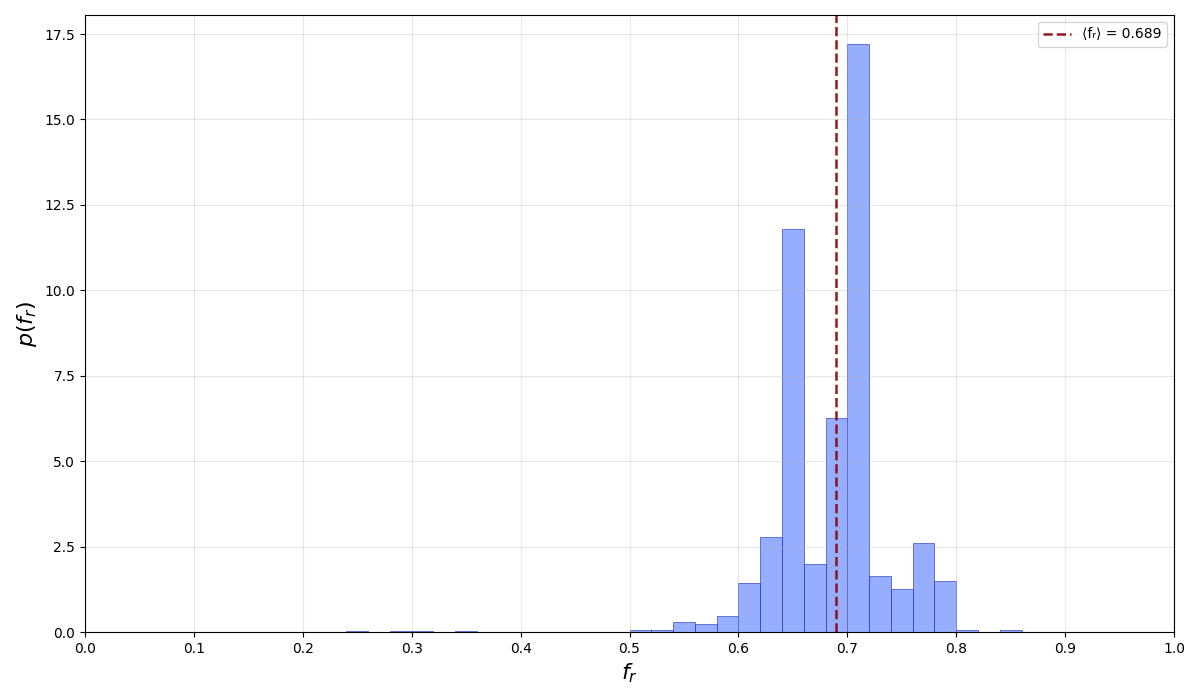}
  \includegraphics[width=0.85\columnwidth]{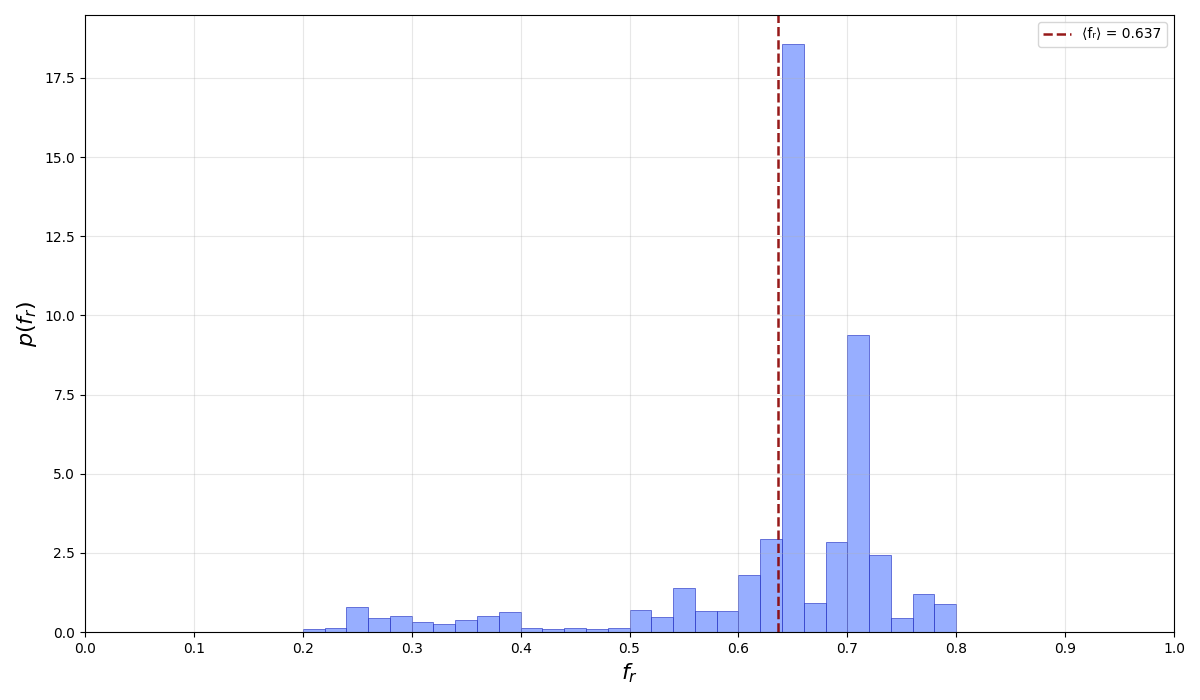}
\end{center}
\caption{\label{fig6} Same as Fig.~\ref{fig2} but for the GINOF model
  with the opinion conviction threshold
  $Z_c=0.1$ at  $W=0.05$ (top); $0.015$ (middle); $0.005$ (bottom), 
  and respectively $<f_r> = 0.540; 0.689; 0.637$ from top to bottom;
  here $N_r=10^5$.
}
\end{figure}

At first glance, it seems that it is sufficient
to consider the GINOF model
with the opinion conviction threshold $Z_c=0$
taking a certain moderate value of
vote amplitude influence $W$. However,
in the framework of GINOF at $Z_c=0$ even a very small value
$W=0.005$ produces a complete change
of the probability distribution
compared to the INOF case with white nodes
or the GINOF case at $Z_c=0, W=0$
(see Fig.~\ref{fig5} and Fig.~\ref{fig2}).
The reason for this drastic change in the distributions
is that at $Z_c=0$, even a very small value of $W \ll 1$
leads to the process where the crowd electors
easily convince their neighbors to adopt a red or blue opinion 
that rapidly increase their vote amplitude influence
up to $W=1$; subsequently, the elite influence
becomes  weak and $f_r$ values are distributed
around $f_r \approx 0.5$, corresponding
to the initial fractions of red and blue
opinions of non-fixed nodes (see Fig.~\ref{fig5}).
In this Fig.~\ref{fig5}, the elite influence
is still present with $<f_r> = 0.575$
but we see that even such a small value as $W=0.005$
gives a qualitative change of the probability distribution
$p(f_r)$ of Fig.~\ref{fig2}.

Thus, it is more adequate to introduce the opinion conviction
threshold $Z_c >0$ as described in Section \ref{sec2}.
We choose $Z_c =0.1$ so that it is close to the minimum value $a_{min} = 0.125$
of the matrix elements of the weighted adjacency matrix $A_{ij}$ (excluding zero elements).
The evolution of the probability distribution
with an increase in the vote amplitude influence $W$
is shown in Fig.~\ref{fig6}. For small $W \leq 0.005$,
the initial distribution $p(f_r)$ at Fig.~\ref{fig2}
remains practically unchanged; then, with an increase to $W=0.015$,
it starts to be modified, and at $W=0.05$, the initial structure
of Fig.~\ref{fig2} is completely washed out, with
$p(f_r)$ being close to that of Fig.~\ref{fig5}.

The results of Fig.~\ref{fig7} are
obtained for one specific initial random configuration
of up-down spins of non-fixed nodes, but
we have verified that the same results hold
for other random configurations.

\subsection{Phase transition of opinion formation}

\begin{figure}
\begin{center}
\includegraphics[width=0.99\columnwidth]{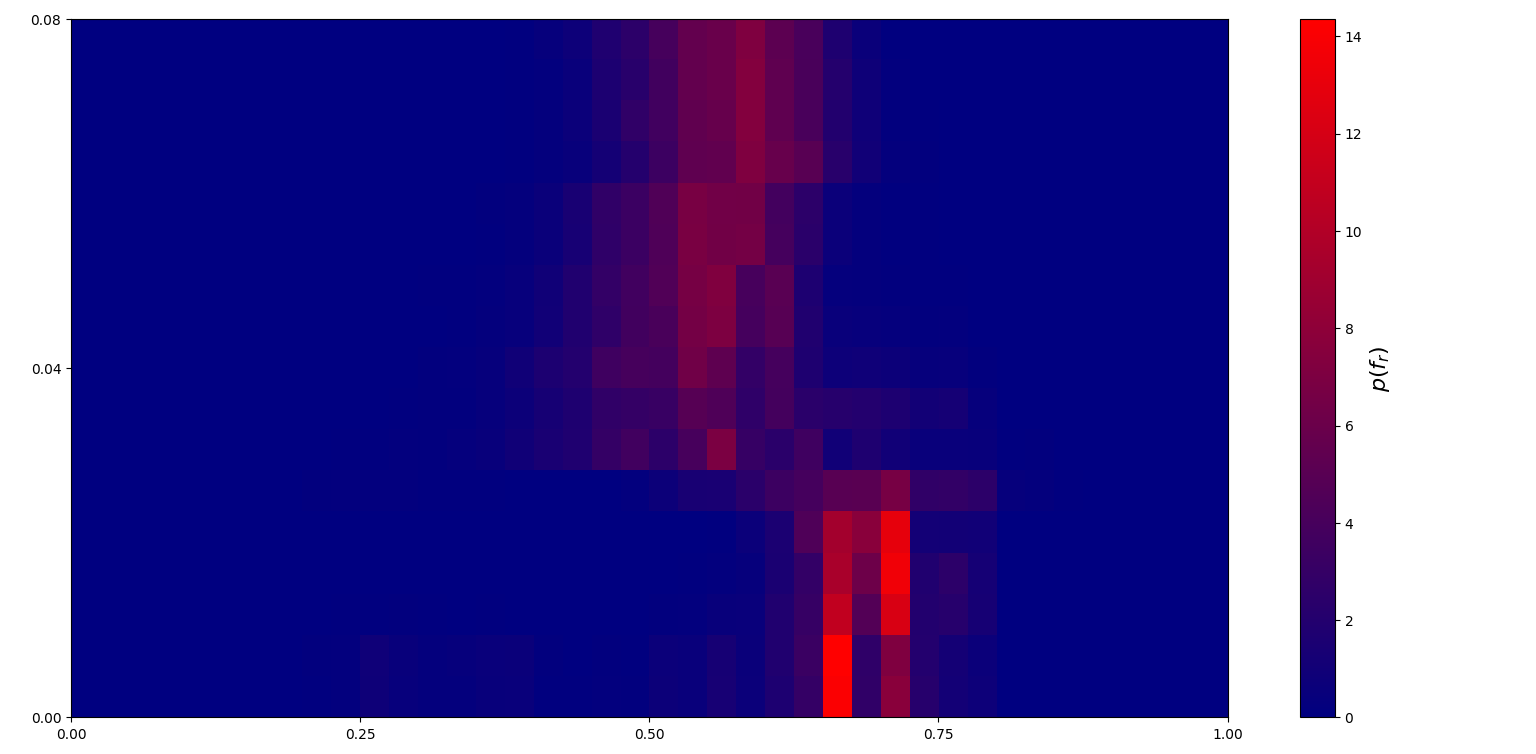}%
\end{center}
\caption{\label{fig7}
   Probability of red nodes $p(f_r)$ shown by color 
   as the function of $x=f_r$ (taken for 40 columns in the range
   $0 \leq f_r \leq1$) and $y=W$ (taken for 17 $W$ equidistant values
   in the range $0 \leq W \leq  0.08$) for the case with
   the opinion conviction threshold $Z_c=0.1$ in the GINOF model
   (there are in total
   $N_{cell} = 680$ cells). Data are obtained with
   $N_r=10^4$ pathway realizations for each $W$ value.}
\end{figure}

The results of Fig.~\ref{fig6} indicate that
there is a phase transition from the regime
at $W < W_{cr}$,
where the elite imposes its opinion,
to a regime at $W>W_{cr}$ where the elite influence is weak
and the elections are mainly affected by
votes from crowd electors. This transition
is illustrated in Fig.~\ref{fig7},
which gives the critical vote amplitude influence
$W_{cr} \approx 0.022$. 
We argue that this critical $W_{cr}$ value is determined
by the condition that the votes of all neighbors
can exceed the opinion conviction threshold so that
\begin{equation}
\label{eqwcr}
W_{cr} \approx Z_c/\kappa .
\end{equation}
In our case, the average number of neighbors is
$\kappa = N_{\ell}/N \approx 4.8$ so that
for $Z_c=0.1$, which gives $W_{cr} \approx 0.021$,
which is close to the above numerical value of Fig.~\ref{fig7}.
It is possible that for networks with a high
number of links per node $\kappa \gg 1$ a more accurate
estimate may be required.

Thus, the obtained results for the GINOF model
demonstrate that in the presence of an opinion conviction threshold,
the elections on social networks are characterized 
by a transition from a phase
where elections are dominated by the elite opinion
to a phase dominated by the votes of crowd electors.
This transition takes place when
the vote amplitude influence $W$ exceeds the
critical value $W_{cr}$ given by the relation (\ref{eqwcr}).

\section{Discussion}
\label{sec4}

In this work, we have generalized the model of opinion
formation on directed Ising networks (INOF) introduced
in \cite{inof,ising25}. This generalized GINOF model
is applied to an undirected social network of scientific collaboration
studied by Newman in \cite{newman2001,newman2006,newmannets,newman2006ref84}.
The new elements of the GINOF model compared to the INOF one
are as follows: in addition to fixed-opinion nodes,
considered as the society's elite,
all non-fixed nodes are initialized with random opinions—half red and half blue.
Furthermore, these non-fixed nodes initially
have a weak amplitude influence ($W \ll 1$), which self-consistently increases
during the asynchronous Monte Carlo process that simulates
an election campaign. In addition, any change of opinion of a given
spin node (a spin flip) takes place only if the
modulus of the majority score of a given node's neighbors' opinions
is above a certain opinion conviction threshold.

We show that for the GINOF model
of elections on undirected social networks
there is a phase transition from
elections dominated by the elite opinion
to a phase where the elite cannot affect the elections
and the vote results are determined by opinions of electors.
We also demonstrate that the Erd\"os barrage
can significantly affect the probability distribution
of red and blue nodes.

At present, there are numerous undirected networks
functioning in human society and various
scientific fields, such as Facebook \cite{facebook}, VK \cite{vk}
and the protein-protein interaction network STRING \cite{string}.
We hope that the GINOF model will find
useful applications in these domains.

\bigskip
\noindent {\bf Acknowledgments}
We thank K.M. Frahm (LPT) for  useful discussions.
This work has been partially supported through the grant
NANOX $N^o$ ANR-17-EURE-0009 in the framework of 
the Programme Investissements d'Avenir (project MTDINA).
\bigskip

\noindent {\bf Data Availability Statement} This manuscript has no
associated data or the data will not be deposited. [Author's
comment: There are no external data associated with the
manuscript.]
\bigskip

\noindent {\bf Author contribution statement} All authors contributed equally to all stages of this work.
  
\bigskip

\bigskip

%%% Multicolumn figures are declared using the corresponding starred figure
%%% environment:
%x\begin{figure*}
%%% Use this one if you have a page-wide figure to typeset the caption in two 
%%% columns:
%   \twocolcaption
%%% Use this one if you have a substantially less than page-wide figure to 
%%% typeset the caption to the right of the figure:
%   \sidecaption
%x  \includegraphics[width=\textwidth]{fig1.eps}%
%x  \caption{Fig.1 caption \label{fig1} 
%x  }
%x\end{figure*}

%%% Use the following two code lines if you wish to generate your bibliography with BibTeX;
%%% please replace first the string "demo" below with the name(s) of
%%% the BibTeX data base(s) you want to use.
%%% The resulting bibliography-output (the contents of the .bbl file)
%%% must be pasted into this file before submission.
%%% 
%%% \bibliographystyle{andp2012}
%%% \bibliography{demo}

\begin{thebibliography}{00}
%%% The number of zeroes here should correspond to the number of digits in the
%%% number of bibliography entries here: if there are up to 9 entries, put one
%%% zero; if there 10 up to 99 entries, put two zeroes; and so on.

%\bibitem{xxx}
%  * On stage from Lyc\'ee general et technologique Alphonse Daudet, Nimes, France

\bibitem{fortunato09} C.~Castellano, S.~Fortunato, and V.~Loreto,
  {\it  Statistical physics of social dynamics},
  Rev. Mod. Phys. {\bf 81},~591 (2009).
\bibitem{dorogovtsev10}  S.~Dorogovtsev,
  {\it  Lectures in Complex Networks},  Oxford University Press,  Oxford, UK (2010).
\bibitem{newmanbook} M.~Newman, {\it Networks},
        Oxford University Press, Oxford, UK (2018).
\bibitem{soc1} Wikipedia contributors, {\it  Social media use in politics},
  Wikipedia, The Free Encyclopedia,
  \url{https://en.wikipedia.org/wiki/Social_media_use_in_politics}
  (Accessed 24 October 2025).
\bibitem{soc2} T.~Fujiwara, K.~Muller,  and C.~Schwarz,
  {\it The Effect of Social Media on Elections: Evidence from The United States},
  J. Eur. Economi Ass., jvad058 (2023);
  \url{https://doi.org/10.1093/jeea/jvad058},
\bibitem{galam82} S.~Galam, Y.~Gefen, and Y.~Shapi,
  {\it Statistical physics of social dynamics},
  Journal of Mathematical Sociology {\bf 9(1)},~1 (1982);
  \url{https://www.tandfonline.com/doi/abs/10.1080/0022250X.1982.9989929}.
\bibitem{galam86} S.~Galam,
  {\it Majority rule, hierarchical structures, and democratic
  totalitarianism: A statistical approach},
  Journal of Mathematical Psychology {\bf 30}, 426 (1986);
  \url{https://doi.org/https://doi.org/10.1016/0022-2496(86)90019-2}.
\bibitem{sznajd00} K.~Sznajd-Weron, and J.~Sznajd,
  {\it Opinion evolution in closed community},
  International Journal of Modern Physics C {\bf 11}, 1157 (2000);
  \url{https://doi.org/10.1142/S0129183100000936}.
\bibitem{sood05} V.~Sood, and S.~Redner,
  {\it Voter Model on Heterogeneous Graphs},
  Phys. Rev. Lett. {\bf 94},~178701 (2005);
  \url{https://doi.org/10.1103/PhysRevLett.94.178701}
\bibitem{watts07} D.J.~Watts, and P.S.~Dodds,
  {\it Influentials, Networks, and Public Opinion Formation},
  Journal of Consumer Research {\bf 34}, 441 (2007);
  \url{https://doi.org/10.1086/518527}.
\bibitem{galam08} S.~Galam,
  {\it Sociophysics: a review of Galam models},
  International Journal of Modern Physics C {\bf 19},~409 (2008);
  \url{https://doi.org/10.1142/S0129183108012297}.
\bibitem{kandiah12} V.~Kandiah, and D.L.Shepelyansky,
  {\it PageRank model of opinion formation on social networks},
  Physica A {\bf 391}, 5779 (2012);
  \url{https://doi.org/10.1016/j.physa.2012.06.047}.
\bibitem{eom15} Y.H.~Eom, and D.L.~Shepelyansky,
  {\it  Opinion formation driven by PageRank
    node influence on directed networks},
  Physica A {\bf 436}, 707 (2015);
  \url{https://doi.org/https://doi.org/10.1016/j.physa.2015.05.095}.
\bibitem{memory1} J.J.~Hopfield, {\it Neural networks and physical systems with
  emergent collective computational abilities},
  Proc. Nat. Acad. Sci. {\bf 79(8)}, 2554 (1982);
  \url{https://doi.org/10.1073/pnas.79.8.2554}.
\bibitem{memory2}  M.~Benedetti, L.~Carillo, E.~Marinari, and M.~Mezard,
  {\it Eigenvector dreaming},
  J. Stat. Mech.  013302 (2024);
  \url{https://doi.org/10.1088/1742-5468/ad138e}.
\bibitem{brics} C.~Coquide, J.~Lages, and D.L.~Shepelyansky,
  {\it Prospects of BRICS currency dominance in international trade},
  Appl. Netw. Sci. {\bf 8}, 65 (2023);
  \url{https://doi.org/10.1007/s41109-023-00590-3}.
\bibitem{inof} L.~Ermann, and D.L.~Shepelyansky,
  {\it Confrontation of Capitalism and Socialism in Wikipedia Networks},
  Information {\bf 15}, 571 (2024);
  \url{https://www.mdpi.com/2078-2489/15/9/571}.
\bibitem{ising25}  L.~Ermann, K.M.~Frahm, and D.L.~Shepelyansky,
  {\it Opinion formation in Wikipedia Ising networks},
  Information {\bf 16}, 782 (2025);
  \url{https://www.mdpi.com/2078-2489/16/9/782}.
\bibitem{dorogovtsev} S.N.~Dorogovtsev, A.V.~Goltsev, and  F.F.~Mendes,
  {\it Ising model on networks with an arbitrary distribution of connections},
  Phys. Rev. E {\bf 66},  016104 (2002),
  \url{https://doi.org/10.1103/PhysRevE.66.016104}.
\bibitem{bianconi} G.~Bianconi,
  {\it  Mean field solution of the Ising model on a Barabási–Albert network},
  Phys. Lett. A {\bf 303}, 166 (2002),
  \url{https://doi.org/10.1073/pnas.79.8.2554}.
\bibitem{newman2001} M. E. J. Newman, 
  {\it Scientific collaboration networks. II. Shortest paths, weighted networks, and centrality}, 
  Phys. Rev. E {\bf 64}, 016132 (2001).
\bibitem{newman2006} M. E. J. Newman, 
  {\it Finding community structure in networks using the eigenvectors of matrices}, 
  Phys. Rev. E {\bf 74}, 036104 (2006).
\bibitem{newmannets} M.~E.~J.~Newman, {\it Network data},
       \url{http://www.umich.edu/~mejn/netdata},
         (Accessed 25 October 2025).
\bibitem{newman2006ref84} M. E. J. Newman, {\it Community Centrality},
  \url{http://www.umich.edu/~mejn/centrality}
  (Accessed 25 October 2025). 
  \bibitem{ourwebpage}% 
  K.Bukina and D.L.~Shepelyansky,
   {\it GINOF data sets}  \url{https://www.quantware.ups-tlse.fr/QWLIB/GINOF4socialnets/}
   (Accessed 7 November 2025). 
\bibitem{fpu70} K.M.~Frahm, and D.L.Shepelyansky,
  {\it Wealth thermalization hypothesis and social networks},
  arXiv:2506.17720 [cond-mat.stat-mech] (2025).
\bibitem{meyer} A.~M.~Langville, and C.~D.~Meyer, {\it Google's PageRank
         and Beyond: The Science of Search Engine Rankings}, 
         Princeton University Press, Princeton (2006).
\bibitem{albert}  R.~Albert, and J.~Thakar,
        {\it Boolean modeling: a logic-based dynamic approach
         for understanding signaling and regulatory networks
         and for making useful predictions},
          WIREs Syst. Biol. Med. {\bf 6}, 353 (2014),
          \url{https://wires.onlinelibrary.wiley.com/doi/10.1002/wsbm.1273}.
\bibitem{levine} S.~Tripathi, D.A.~Kessler, and  H.~Levine, 
  {\it Biological Networks Regulating Cell Fate Choice Are Minimally Frustrated},
  Phys. Rev. Lett. {\bf 125},~088101 (2020),
  \url{https://wires.onlinelibrary.wiley.com/doi/10.1002/wsbm.1273}.
\bibitem{fibrosis2} K.M.~Frahm, E.~Kotelnikova, O.~Kunduzova, and D.L.~Shepelyansky,
  {\it Fibroblast-Specific Protein-Protein Interactions for Myocardial
    Fibrosis from MetaCore Network}, Biomolecules {\bf 14}, 1395 (2024),
  \url{https://www.mdpi.com/2218-273X/14/11/1395}.
\bibitem{facebook} Facebook \url{https://www.facebook.com/}
  (Accessed 7 November 2025).
\bibitem{vk} VK \url{vk.com},
  (Accessed 7 November 2025).
\bibitem{string} STRING \url{https://string-db.org/},
  (Accessed 7 November 2025).
  
\end{thebibliography}
%%% 
%%% If you are doing it by hand make sure to put the correct make up into 
%%% it.  Please see below for the usual keys; and please keep in mind that all the custom 
%%% macros (\jr, \othercit} and remarkably also \textsc  here have no effect 
%%% regarding typesetting.  They are crucial for the hypertext markup of the 
%%% data, though.
%%% 
%%% The macros are:
%%% \textsc for authors' names;  also for editors' names, if there are no authors;
%%% \jr for (abbreviated) journal names;
%%% \othercit to be used as a prefix to \bibitem in non-journal entries and like 
%%% a common makro for partial entries in multi-entry \bibref-constructs.
%%% 
%%% Replace the following example bibliography with your references
%%% before submission:

\noindent {\bf NOTE(*)} On stage from Lyc\'ee general et technologique Alphonse Daudet, Nimes, France

\end{document}